# A Critique of General Relativity

*Wasley S. Krogdahl*
*Professor Emeritus, Astronomy and Physics*
*University of Kentucky*

ABSTRACT
*General relativity's successes and limitations are compared to those of special relativity.*

General relativity has become the orthodox formulation of gravitational theory. It is unquestioned in its several applications in astronomy and physics. Its preeminence and its implications were accepted serially following its promulgation by Albert Einstein in 1915, but its accumulating apparent successes have won it progressive dominance, there being virtually no challenge today to its correctness. It is the purpose of this essay to examine this status.

General relativity's successive triumphs included (a) accounting for a previously well known but unaccounted for advance of the perihelion of Mercury, (b) the re-fraction of light at the limb of the sun and the gravitational lensing of distant galaxies, (c) the echo delay of a sun-grazing radio signal, (d) gravitational redshift, (e) radiation by gravitationally accelerated bodies[1], and (f) other effects, mostly subtle.

Newtonian gravitational theory has been unsuccessful in accounting correctly (if at all) for any of these effects, though more than adequate for most purposes. The special theory of relativity (1905), successfully treated electromagnetism and mechanics of inertial systems but appeared to be incapable of treating gravitation.

With this background, general relativity's successes have given it a monopoly on all matters concerning gravitation. To question general relativity, therefore, verges on heresy, and those who raise doubts concerning it may reap indifference if not the scorn reserved for heretics and cranks. Nevertheless, every theory, including general relativity, should be fair game for critical analysis. Without further apology, let us consider some of the shortcomings or outright failures of the theory.

The last major implication of general relativity was the prediction of "black holes". These are implied by the Schwarzschild metric[2], which has a singular surface (event horizon) at a specified distance about a compact mass. Point singularities abound in physics, but a black hole is unique. This of itself should be grounds for caution. Furthermore, there is no unambiguous test for the observational identification of a presumptive black hole. It cannot be discriminated from a possible "non-black hole" of equal mass and radius. One may therefore be forgiven a healthy skepticism in the confident identification of a multitude of these

objects; their implied existence should be considered a possible failure of the theory.[3]

This is not the only caveat. Photons presumably cannot escape from a black hole.
What happens, therefore, to a photon emitted radially within a black hole? It cannot decelerate and reverse itself, as would a mass particle. It must be redshifted to extinction. This would violate the equation for gravitational redshift.[4]

Some of the earliest applications of general relativity were to models of the universe. It is characteristic of all such models that they are finite in mass, volume and age.[5] As was pointed out by E. A. Milne in the 1930s, they must therefore have unique mass and velocity centroids, features which relativity was presumably intended to avoid. Milne also showed that expanding models required that matter be created at the boundary during expansion and that oscillating models required the destruction of matter during the contracting phase. These models therefore lacked cosmic background radiation, though this was not remarked at the time since the cosmic background radiation would not be discovered until several decades later.

A consequential objection to all general relativistic models of the universe is that they are hydrodynamic. That is, they postulate that the matter of the universe is spread continuously throughout. The real universe, however, is atomistic and granular. It consists of discrete objects from electrons and protons to molecules, planets, stars, galaxies and clusters of galaxies. General relativity is an essentially field theory, not capable of accounting for the granular appearance of the cosmic background radiation or the vast vacancies between galaxies.[6]

The theory is also ambiguous in several respects. For one, it cannot say whether its model universes are oscillating, static, or forever expanding. It depends upon observation to identify which kind of universe is the actual one. This is not a telling objection, since appeals to observation are always in order for any theory. One may view such uncertainty, however, as a comparative disadvantage with respect to a theory whose model(s) is (are) specific.

Related to this ambiguity is the value of the "cosmological constant". It could conceivably be positive, negative or identically zero (non-existent). A complete theory would ideally be able to give a reason for a particular choice. The cosmologic-al constant is said to be required in order to account for a presumed acceleration of the expansion of the universe. However, this hypothetical acceleration is rather a consequence of neglecting to take proper account of time dilation in clocks receding at high velocity. The acceleration is not real. Moreover, a positive value would imply a force of gravitation between two masses which, beyond a certain point, increases with distance; this is a highly counterintuitive result.[7]

The general theory is mute concerning the controversy over the value of the Hubble "constant". Kinematic relativity, a Lorentz invariant theory, resolves the controversy by showing that both sides are correct. The Hubble "constant" is not constant but increases with distance, as the observations show.[8]

General relativity has nothing to say concerning "dark matter" or "dark energy", currently described as "mysterious" and requiring much further research. The presence of "dark matter" was first implied by observations of the motions of stars at the outskirts of galaxies; they revolved about their respective galactic nuclei at velocities far greater than could be justified by the amount of visible matter in those galaxies. The seeming discrepancy is great; "dark matter" would have to amount to some six times that of visible or "baryonic" matter (the sum of the masses of all the protons, neutrons, electrons and other fundamental particles).

If the general theory is defective in the particulars cited above, by what should it be replaced? Unless there is a more successful alternative, a catalogue of its failures or limitations warrants only a greater effort to modify or extend the theory to correct its shortcomings. It is perhaps ironic that the shortcomings of the theory of general relativity can be remedied by appeal to the theory of special relativity.

The special theory (also known as "Lorentz invariant relativity") was dismissed because it seemed unable to predict the several relativistic effects such as those which established the dominance of the general theory. It is now clear that this is not a valid objection; when the equations of gravitation are correctly formulated, the special theory can predict all the same effects as well as offer a model universe which is not objectionable on the grounds which disqualify the general theory. What is needed is not a new law but a new expression for the mass which incorporates the mass-energy relation $m = E/c^2$, the same relation which underlies the destructive power of the atom bomb and is the source of the sun's radiation.

Lorentz invariant cosmology holds promise of being able to account for the ratio of gravitational mass of galaxies to their baryonic masses (though this requires a tedious computation yet to be accomplished); i.e., it conceivably could account for the existence of so-called "dark matter"[9]. General relativity does not. Given the observationally determined ratio of 6:1, Lorentz invariant theory can then predict correctly the value of "dark energy". General relativity cannot. "Dark matter" is, simply, the apparent increase of mass induced in gravitating masses by the presence of other nearby masses; it is a minute amount except when the number of masses is very large and their separations small. It is never apparent in intergalactic regions, for example. "Dark energy" is the greater kinetic energy implied by the conservation of energy and the greater gravitational potential of the "dark matter". No new "mysterious" particles are needed.

The mystique of general relativity and "space-time curvature" has proven to be extremely powerful. At least some of its hold must reside in the very mystery evoked by the term "space-time curvature" (often mistakenly referred to as "the curvature of space" in popular literature). However, the mystery may be effectively dispelled by noting that the phrase is merely equivalent to the term "gravitational acceleration". If there is no space-time curvature, there is no gravitational acceleration; if there is no gravitational acceleration, there is no space-time curvature.

2.

3.

4.

5.

6.

7.

8.

9.